# Smartening the Environment using Wireless Sensor Networks in a Developing Country

Al-Sakib Khan Pathan
*Department of Computer Engg.*
*Kyung Hee University, Korea*
spathan@networking.khu.ac.kr

Choong Seon Hong
*Department of Computer Engg.*
*Kyung Hee University, Korea*
cshong@khu.ac.kr

Hyung-Woo Lee
*Department of Software*
*Hanshin University, Korea*
hwlee@hs.ac.kr

*Abstract* — The miniaturization process of various sensing devices has become a reality by enormous research and advancements accomplished in Micro Electro-Mechanical Systems (MEMS) and Very Large Scale Integration (VLSI) lithography. Regardless of such extensive efforts in optimizing the hardware, algorithm, and protocols for networking, there still remains a lot of scope to explore how these innovations can all be tied together to design Wireless Sensor Networks (WSN) for smartening the surrounding environment for some practical purposes. In this paper we explore the prospects of wireless sensor networks and propose a design level framework for developing a smart environment using WSNs, which could be beneficial for a developing country like Bangladesh. In connection to this, we also discuss the major aspects of wireless sensor networks.

## 1. Introduction

The notion of smart environment is becoming a reality with the advancements of various smart technologies. Smart environments represent the future evolutionary development step for the real world environment of present time. A smart environment, like any conscious organism, relies first and foremost on sensory data acquired from multiple sensors in distributed locations of real world. It gathers information about its surroundings as well as about its internal workings [1]. In the recent years, an exciting new type of networks has emerged, called Wireless Sensor Networks (WSN) [2], [3]. The deployment of such networks not only effectively acquires the data from different locations and then distribute to the management centers but also facilitates other applications for facing disasters and other environmental issues. In this paper, we explore the scope to deploy Wireless Sensor Networks for developing a smart environment especially in Bangladesh to facilitate various sophisticated systems to face disasters like flood, tsunami, and cyclones as well as to enhance road traffic monitoring system. In fact, the notion of smart environment has a great potential for a developing country like Bangladesh which faces different types of natural disasters each year. While some other works [4], [5], [6] focus on specific topics like smart homes, smart classrooms etc. as a part of smart environment, we explore the promise of wireless sensor networks for smartening the environment by up-gradation of various monitoring and warning systems aided with wireless sensing technology.

The organization of this paper is as follows: Section 2 discusses some related works on smart environment, Section 3 gives an overview of smart sensors and smart wireless sensor networks, Section 4 mentions some positive aspects of WSN that makes it attractive for its deployment in real-world, Section 5 tells about the application view of smart environment and systems in Bangladesh using WSNs and Section 6 concludes the paper.

## 2. Related Works

Smart sensor nodes show great promise for making our lifestyle better than that of today. The smartness and type of tasks expected from smart sensors consequently led to the notion of smart environment. In some previous studies smart sensors and their possible applications were examined. Henderson et. al. [7] studied the sensors and proposed deployment and exploitation of large number of sensors to obtain a trigger action over a large geographic area. They discussed several issues in their work including sensor distribution pattern and local sensor frames. Hamrita et. al. [8] gives an overview of smart dust technology and illustrates the application view using their prototype which uses the Berkeley smart sensor "Mote". Zhang et. al. [9], in their study explored the smart sensors and smart sensor networks. Their work discussed some issues related to smart sensor and smart sensor network that are to be resolved. [10] presents an overview of ad-hoc smart environments and mentions a number of research issues and challenges for designing such environments. Some other studies [11], [12] investigated various facets of the research on smart environment using smart sensing technologies.

In this paper, we investigate how smart sensors could be used for developing a smart environment using wireless sensor network technologies in an actual setting. For our study we considered a developing country Bangladesh and analyzing various aspects we propose a design level framework towards achieving smart environment.

## 3. Smart Sensors and Wireless Sensor Network – A Background

This work was supported by MIC and ITRC Project

Massive advancements in wireless communications, Micro-Electro-Mechanical Systems (MEMS), and optics have opened the new chapter of modern civilization, populated with small, low-power, cost-effective, autonomous devices, termed sensor nodes, which would pervade our society redefining the way it is at present [13]. Sensor nodes are of the combination of sensing and special-purpose computing devices tied with wireless communications. When networked, such sensor nodes would build up the part of larger systems, providing data, as well as performing and controlling multitude of tasks and functions. Small size and cost of individual sensor nodes would be the key ingredient for a large number of applications both in ordinary as well as harsh environments. Given the utility of sensor networks in environmental data collection, surveillance, and target tracking, they can aid numerous applications as their requirements vary along with the time-space-context continuum [14]. Sensor networks can be used in support of preparation and prevention during various phases of pre-event, rapid response during the event, and post recovery along with analysis after the event. To benefit the environment, in practical, these large number of miniaturized commodity sensor nodes could be installed, for example in buildings, on roads, in vehicles, at the riverbanks, or at coastal areas etc. Deployment of new sensor nodes may take place on demand at any time at designated locations, referred to as area of interest (AOI) or at random in specified areas.

A smart sensor node is a combination of sensing, processing and communication technologies. Figure 1 shows the basic architectural components of a sensor node. The sensing unit senses the change of parameters, signal conditioning circuitry prepares the electrical signals to convert to the digital domain, the sensed analog signal is converted and is used as the input to the application algorithms or processing unit, the memory helps processing of tasks and the transceiver is used for communicating with other sensors or the base stations or sinks in WSN.

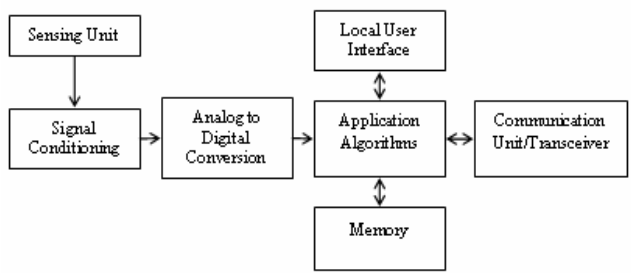

**Figure 1: Basic architectural components of a smart sensor**

Sensors can monitor temperature, pressure, humidity, soil makeup, vehicular movement, noise levels, lighting conditions, the presence or absence of certain kinds of objects or substances, mechanical stress levels on attached objects, and other properties [15]. Their mechanism may be seismic, magnetic, thermal, visual, infrared, acoustic, or radar [16]. A smart sensor is also capable of self-identification and self-diagnosis. The mechanisms of smart sensors work in one of three ways: by a line of sight to the target (such as visual sensors), by proximity to target (such as seismic sensors), and by propagation like a wave with possible bending (such as acoustic sensors).

Sensor networks are predominantly data-centric rather than address-centric. In such a network, queries are directed to a region containing a cluster of sensor nodes rather than specific sensor addresses. Given the similarity in the data obtained by sensors in a dense cluster, aggregation of the data is performed locally. That is, a summary or analysis of the local data is prepared by an aggregator node within the cluster, thus reducing the communication bandwidth requirements. Aggregation of data increases the level of accuracy and incorporates data redundancy to compensate node failures. A network hierarchy and clustering of sensor nodes allows for network scalability, robustness, efficient resource utilization and lower power consumption which are some of the key issues in WSN.

## 4. Why Wireless Sensor Networks?

Fundamental objectives of sensor networks are reliability, accuracy, flexibility, cost effectiveness and ease of deployment. Key characteristics and benefits of WSN (Wireless Sensor Networks) are outlined below:

- **Sensing accuracy:** The utilization of a larger number and variety of sensor nodes provides potential for greater accuracy in the information gathered as compared to that obtained from a single sensor.
- **Area coverage:** This implies that fast and efficient sensor network could span a greater geographical area without adverse impact on the overall network cost.
- **Fault tolerance:** Device redundancy and consequently information redundancy can be utilized to ensure a level of fault tolerance in individual sensors.
- **Connectivity:** Multiple sensor networks may be connected through sink nodes, along with existing wired networks (e.g. Internet). The clustering of networks enables each individual network to focus on specific areas or events and share only relevant information.
- **Minimal human interaction:** Having minimum human interaction makes the possibility of having less interruption of the system.
- **Operability in harsh environments:** Sensor nodes, consisting of robust sensor design, integrated with high levels of fault tolerance can be deployed in harsh environments that make the sensor networks more effective.
- **Dynamic sensor scheduling:** Implying some scheduling scheme, sensor network is capable of setting priority for data transmission.

## 5. Application of WSN toward Bangladesh

By the geographical location in the globe, Bangladesh is very susceptible to many environmental calamities like, flood, cyclone, tsunami etc. Good warning systems could effectively help to mitigate the damages caused by these natural disasters.

Hence, the development of wireless sensor networks to assist meteorologists has a great deal of national importance in Bangladesh.

Sensor networks provide the ability to gather accurate and reliable information, to enable early warnings and rapid coordinated responses to potential threats. This encompasses the ability to save lives through environmental monitoring of natural disasters. Only proper infrastructure through long term research and implementation of these technologies can make huge difference in a country like Bangladesh. Environmental sustainability is of great importance in such geographical locations in the world, where it could be improved through sensor monitoring, by protecting valuable resources, and collecting valuable information previously considered too difficult and too costly.

### 5.1. Flood and Water Level Monitoring System

It takes immense supremacy and courage to confront the situation, when natural disasters, such as earthquakes, floods, tsunami etc. occur. In order to cope with such disasters in a fast and highly coordinated manner, the optimal provision of information concerning the situation is an essential pre-requisite. Over the course of the past decade, tremendous changes to the global information and communication infrastructure have taken place, including the popular uptake of the Internet, the staggering growth and plummeting costs of mobile telecommunications, and the implementation of advanced space-based remote sensing and satellite communication systems. Among the new achievements and technologies of global information and communication infrastructure, there emerge new technologies like WSN. Nevertheless, there is a great deal of potential for sensor networks to be deployed for the flood monitoring system, especially in a tropical region like Bangladesh.

Even though, the initial idea for WSN was to deal with such systems, many researchers around the world have come across with various features of WSN but it is yet to setup such systems in developing countries. A framework of such a system is brought by this research work. The following section outlines a system aided with wireless sensor network for flood controlling and warning system in Bangladesh.

Any country that is under the threat of flood requires a flood monitoring, controlling, and warning system. It is important to classify various phases consisting of different levels of activities underlying with such system development. Starting with the first phase of data collection, level one is to deal with the physical deployment of sensing devices in the riverbanks and implementation of an effective localization scheme depending on the situation and environment. The flow path of the river, past records of water flow and future prediction of the route of the river influence the placements of the wireless sensors. The sensors form clusters to communicate with the local base stations. The local base stations are powerful enough to communicate with one another directly using wireless communications. The data sent from the sensors are aggregated in the local base stations to provide as inputs to the data processing center (s). Figure 3 shows a pictorial view of the deployment of sensor nodes and data aggregation.

Level two deals with the setup of local base stations as well as with data communication at district level. Level three could be involved with the central monitoring system at the headquarters to process acquired data. Data analysis then takes place either at headquarter or at outside research centers that particularly do high-risk flood analysis. Once, the processed data is ready with findings having minimal level of errors, next phase starts reversed to the first phase.

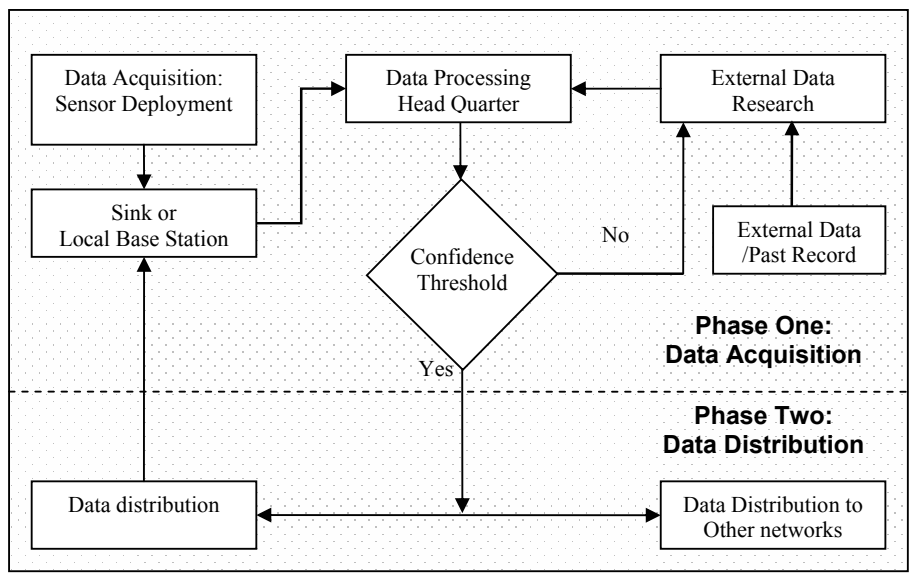

**Figure 2: Smart System aided with Wireless Sensor Network**

Second phase involves data distribution network, where processed data are sent to different factors involved within the network. For instance, a flood warning issued by the headquarter could be sent to local mobile phone companies for delivering SMS (Short Message Service) to their mobile subscribers. Similarly, police department or fire department could be notified for any precaution to be taken immediately by the use of various means of distribution networks. Figure 2 illustrates the flow chart of the proposed framework of the system. Data processing starts after the acquisition part takes place and then goes to a confidence threshold.

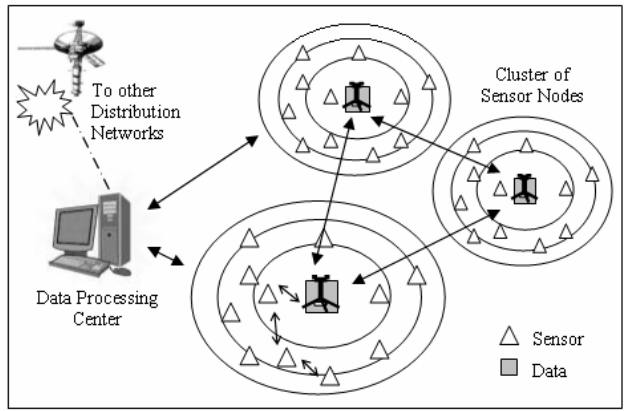

**Figure 3: Data collection and aggregation in wireless sensor network**

Error detection is defined by the predetermined threshold and if necessary sent back to processing center through the External Data Research Center. Once the processed data is ready, second phase starts within distribution network. The overall application of WSN is very demanding in terms of future references as well as for data propagation.

### 5.2. Traffic Monitoring and Controlling

Statistics from the Road Safety Cell (RSC) of the Bangladesh Road Transport Authority (BRTA) show the annual fatality rate in road accidents in Bangladesh is 85.6 per 10,000 vehicles which compares to rates of below 3 per 10,000 vehicles in most developed countries [17]. Another recent study estimates the national cost of such road traffic crashes in Bangladesh at Taka 45 billion (US$ 76 million) which is more than 1.6% of the country's GDP (Gross Domestic Product) [18]. Consequently, the need for having an efficient traffic controlling and monitoring system is very demanding. Deployment of smart sensors along with the roadside made many developed countries possible to collect live data or to monitor irresponsible vehicle violating the speed limit. In case of Bangladesh, placing sensor nodes at the identified spots with the addition of wireless communications could help for the development of a smart traffic monitoring system. Each sensor needs to be placed within predetermined optimum distance ranges, following a special node. This special node would act as local base station equipped with data transmission capabilities to the headquarters. Once the sensed data are received, central station might broadcast the message through radio stations. The overall system is similar to the proposed flood controlling system, except sensor placement locations.

Apart from above mentioned system, a simple traffic signal system could also be equipped with intelligent sensing devices at the road intersections. The smart sensing device would gather information of upcoming objects towards the intersection; perform scheduling to determine the time-to-wait (TTW) interval for signals to be changed. Time-to-wait is determined by the time gap between different crossing objects. Hence, automated signal changing would never keep the motorist waiting in one side for a longer time. Not to mention, TTW could also be human operated for special purposes like medical, military or other emergency situations.

### 5.3. Environmental monitoring

Environmental monitoring system could be crucial in analyzing the important data for forecasting weather and even for measuring the possible environmental threat, that to be prevented in a country like Bangladesh. Many initial WSNs have been deployed for environmental monitoring, which involves collecting readings over time across a volume of space large enough to exhibit significant internal variation [19]. WSN is now also used by the researchers to monitor seabird habitats, for conducting analogous studies of contaminant propagation, building comfort, and intrusion detection. These systems could help implementation of smart environment even in Bangladesh. All these systems require data and their effective and timely processing which is shown earlier in Figure 2.

### 5.4. Future Steps: Effective Localization scheme and Cost Effectiveness of the System

Location awareness of wireless sensor network is among the key elements to be premeditated for the network optimization. Often expensive GPS (Global Positioning System) receivers are being replaced with small low-cost seed nodes that are capable of relaying the location information. In some cases, the AOA (angle of arrival) [20] has the greater importance over the actual location information. Again, some research conducted to find the Virtual Coordination System [21] for Wireless Sensor Networks (WSN) attained greater importance. Other studies have been conducted on Direction-based Localization Scheme [22]. Several studies have been made on the different localization techniques, but none of them considered being the best in all aspects. Hence, it has a great deal of significance and importance to study and compare various challenges concerning various localization schemes for the optimization.

Depending on the system of interests, elevated significance goes to the fact that how effective the system is in terms of cost analysis. Researchers conducting the design and framework, mainly focus on the blueprint of the innovative system they might be working on. Once the design is ready for implementation, it then goes to the industry and MIS specialist in order to find out how successful the new design would be in terms of implementation. This is well experienced in developed countries like United States. However, in a country like Bangladesh representing the third world developing regions of the world, it is often highly demanded to do the cost analysis for implementing any system that is technically sound and sophisticated. Major research work could be carried out in particular to find out cost effectiveness for highly perspective project like wireless sensor networks for flood, cyclone and other natural disaster controlling and warning.

## 6. Conclusion

As technology emerges over the decades, WSN has come to the spotlight for its unattained potential and significance. Consequently, billions of dollars are being committed to the research and development of sensor networks in order to address the operational challenges that are still associated with the large-scale implementation of sensor networks. Without having the proper blueprint, no construction manager could put up any building according to architect's intention. Similar approach applies almost in everywhere for developing any new system in the society. This paper is mainly focused on the design level issues provided with a framework for WSN to be applied on various systems. In case of Bangladesh, flood and water level monitoring, traffic monitoring, and environmental monitoring are among the systems having much potential to be aided with WSN, which would lead the development of a smart environment. Various smart applications and sophisticated systems could share the same sensor nodes, deployed around the particular area of interest (AOI) for performing the job simultaneously. WSN, the emerging technology makes the possibility of recognizing present and predicting future in a way not possible in past. A smart environment is to be regarded as indispensable stage of a real time system for sensing and prevention of any undesirable occurrence.